\documentclass[journal]{IEEEtran}
\usepackage{amsmath}
\usepackage{tikz}
  \usetikzlibrary{arrows,automata,shapes}
\usepackage{algorithm} 
\usepackage[noend]{algpseudocode}

\usepackage{array}
\usepackage{cite}
\usepackage{url}
\allowdisplaybreaks

\newcommand{\eps}{\varepsilon}
\newcommand{\suppc}{\overline{\sup}{\,}}
\newcommand{\infpcCO}{\overline{\inf}{\,\textrm{CO}}}
\newcommand{\infpcC}{\overline{\inf}{\,\textrm{C}}}
\newcommand{\infpcO}{\overline{\inf}{\,\textrm{O}}}

\newcommand{\A}{\mathcal A}
\newcommand{\B}{\mathcal B}

\newtheorem{thm}{Theorem}
\newtheorem{lem}[thm]{Lemma}
\newtheorem{cor}[thm]{Corollary}

\newtheorem{fact}[thm]{Fact}

\begin{document}
\title{Complexity of Infimal Observable Superlanguages\thanks{Supported by the DFG in Emmy Noether grant KR~4381/1-1 (DIAMOND).}}

\author{Tom{\' a}{\v s}~Masopust\thanks{T. Masopust ({\tt masopust{@}math.cas.cz}) is with CFEAD, TU Dresden, Germany, and with Institute of Mathematics, Czech Academy of Sciences
}}

\markboth{}{}

\maketitle

\begin{abstract}
  The infimal prefix-closed, controllable and observable superlanguage plays an essential role in the relationship between controllability, observability and co-observability -- the central notions of supervisory control theory. Existing algorithms for its computation are exponential and it is not known whether a polynomial algorithm exists. In this paper, we study the state complexity of this language. State complexity of a language is the number of states of the minimal DFA for the language. For a language of state complexity $n$, we show that the upper-bound state complexity on the infimal prefix-closed and observable superlanguage is $2^n + 1$ and that this bound is asymptotically tight. It proves that there is no algorithm computing a DFA of the infimal prefix-closed and observable superlanguage in polynomial time. Our construction further shows that such a DFA can be computed in time $O(2^n)$. The construction involves NFAs and a computation of the supremal prefix-closed sublanguage. We study the computation of the supremal prefix-closed sublanguage and show that there is no polynomial-time algorithm that computes an NFA of the supremal prefix-closed sublanguage of a language given as an NFA even if the language is unary. 
\end{abstract}


\section{Introduction}
  \IEEEPARstart{C}{ontrollability} and observability are the central notions of supervisory control theory of discrete event systems in the Ramadge-Wonham framework~\cite{Cieslak88,LinW88,RW89}. They form the necessary and sufficient conditions for the existence of a supervisor that achieves the desired control behavior of a system. 
  In decentralized supervisory control, where more supervisors cooperate to control the system, every supervisor observes and controls part of the system. The observation of a supervisor is modeled by an observation mask or by a natural projection. Cieslak et al.~\cite{Cieslak88} and Rudie and Wonham~\cite{RudieWonham1992} have shown that controllability and co-observability are the central notions in decentralized supervisory control. 
  
  A relationship between controllability, observability and co-observability has been studied by Kumar and Shayman~\cite{KumarS98}, who have shown that the infimal prefix-closed, controllable and observable superlanguage plays the essential role. Another motivation and the importance of infimal superlanguages have been discussed in the fundamental book on supervisory control theory~\cite{CL08}. We have further illustrated its relevance to decentralized supervisory control with communication~\cite{KomendaM15} and to coordination control~\cite{iccse2015}. We refer the reader to these papers for more details and examples.
  
  Infimal superlanguages are of a general interest in supervisory control. There are examples in modular and decentralized control showing evidence that supremal sublanguages do not always suffice to achieve the best (optimal) solution and that the optimal solution may be achieved if infimal superlanguages are involved. The examples show evidence that the combination of supremal sublanguages and infimal superlanguages help achieve optimality if it is not achievable by supremal sublanguages alone~\cite{KomendaM15,iccse2015}. Therefore our interest in infimal prefix-closed, controllable and observable superlanguages. 
  
  Lafortune and Chen~\cite{LC1990} have shown that the infimal prefix-closed and controllable superlanguage can be computed from a deterministic finite automaton (DFA) for the language in linear time. Kumar and Shayman~\cite{KumarS98} have further shown that it is sufficient to consider the computation of the infimal prefix-closed and observable superlanguage of a language $K$ over $\Sigma$ wrt the language $\Sigma^*$. Thus, we focus in this paper on the infimal prefix-closed and observable superlanguage of $K$ wrt $\Sigma^*$ and study its state complexity. 

  {\em State complexity of a language\/} is the number of states of the minimal DFA marking (accepting) the language. Since the minimal DFA is unique (up to isomorphism), state complexity is a complexity measure that is independent of the representation and computation of the language.

  \paragraph*{Our contribution}
  For a language $K$ of state complexity $n$, we show that the upper-bound on the state complexity of the infimal prefix-closed and observable superlanguage of $K$ wrt the language $\Sigma^*$ is $2^n + 1$. We further prove that this bound is asymptotically tight by showing that the worst-case lower-bound state complexity is at least $\frac{3}{4}\cdot 2^{n}-1 = \Omega(2^n)$. Since the state complexity is exponential, so is the time complexity of any algorithm computing the corresponding minimal DFA. 
  
  In addition, our construction shows that a DFA representation of the infimal prefix-closed and observable superlanguage of $K$ wrt the language $\Sigma^*$ can be computed in time $O(2^n)$. 
  
  Our construction involves nondeterministic finite automata (NFAs) and is based on a formula equivalent to the formulae of Rudie and Wonham~\cite{RudieWonham1990} and of Kumar and Shayman~\cite{KumarS98}. The formulae include a computation of the supremal prefix-closed sublanguage. We study the computation of the supremal prefix-closed sublanguage and show that there is no polynomial-time algorithm computing an NFA representation of the supremal prefix-closed sublanguage of a language given as an NFA even if the language is unary.

\section{Preliminaries}
  We assume that the reader is familiar with supervisory control theory~\cite{CL08} and automata theory~\cite{HopcroftU79,sipser}. For undefined notions, the reader is refer to these references.

  The prefix closure of a language $L$ is the set $\overline{L}=\{w\in \Sigma^* \mid \text{there is } u \in\Sigma^* \text{ s.t. } wu\in L\}$; $L$ is {\em prefix-closed\/} if $L=\overline{L}$. The {\em right quotient\/} of a language $L$ wrt a language $M$ is the set $L/M = \{w\in\Sigma^* \mid \text{there is } x\in M \text{ s.t. } wx\in L\}$. If $M=\{a\}$ is a singleton, we simply write $L/a =\{w \in \Sigma^* \mid wa \in L\}$.
  The {\em empty string\/} is denoted by $\varepsilon$. 

  A {\em nondeterministic finite automaton\/} (NFA) is a quintuple $\A = (Q,\Sigma,\delta,Q_0,F)$, where $Q$ is a finite nonempty set of states, $\Sigma$ is an input alphabet, $Q_0\subseteq Q$ is a set of initial states, $F \subseteq Q$ is a set of marked states, and $\delta \colon Q\times(\Sigma\cup\{\eps\}) \to 2^Q$ is a transition function that is extended to $2^Q\times\Sigma^*$ by induction. The language {\em generated\/} by $\A$ is the set $L(\A) = \{w\in \Sigma^* \mid \delta(Q_0,w)\neq\emptyset\}$ and the language {\em marked\/} by $\A$ is the set $L_m(\A) = \{w\in \Sigma^* \mid \delta(Q_0,w)\cap F \neq\emptyset\}$.

  The NFA $\A$ is an (incomplete) {\em deterministic finite automaton} (DFA) if $|Q_0|\le 1$ and $|\delta(q,a)|\le 1$ for every $q \in Q$ and $a \in \Sigma$. Moreover, DFAs do not admit $\eps$-transitions, that is, $\delta$ is a partial transition function from $Q\times \Sigma$ to $Q$. 

  For every NFA $\A$ there exists a DFA $\B$ such that $L_m(\B)=L_m(\A)$ and $L(\B)=L(\A)$. The DFA $\B$ is constructed by the standard {\em subset construction\/}~\cite{sipser} and is called the {\em subset automaton\/} of $\A$. Specifically, for $\A=(Q,\Sigma,\delta,Q_0,F)$, $\B = (2^Q,\Sigma,\delta',Q_0,F')$, where $\delta' \colon 2^Q \times \Sigma \to 2^Q$ is defined as $\delta'(X,a) = \delta(X,a)$ and $F' = \{ R\subseteq Q \mid R\cap F\neq\emptyset\}$.

  Let $\Sigma$ and $\Delta$ be alphabets. An {\em (observation) mask\/} is a map $P\colon \Sigma \to \Delta\cup\{\eps\}$ that is extended to $\Sigma^*$ so that $P(\eps)=\eps$ and $P(sa)=P(s)P(a)$ for $s\in\Sigma^*$ and $a\in\Sigma$. If $L$ is a regular language, then $P(L) = \cup_{w\in L}\, P(w)$ is regular~\cite{ginsburg75}. A mask $P$ is a {\em (natural) projection\/} if $\Delta \subseteq \Sigma$ and $P(a)=a$, for $a\in\Delta$, and $P(a)=\eps$ otherwise.
  The {\em inverse image\/} of a mask $P$, denoted by $P^{-1}\colon 2^{\Delta^*}\to 2^{\Sigma^*}$, is defined as $P^{-1}(L)=\{w \in \Sigma^* \mid P(w)\in L \}$. Regular languages are closed under the inverse image of a mask~\cite{ginsburg75}.

  In the rest, the term {\em language\/} stands for a regular language.

\section{Known and Preliminary Results}
  Let $\infpcCO(K,L(G),\Sigma_u,P)$ denote the infimal superlanguage of $K$ that is prefix-closed, controllable and observable wrt $L(G)$, uncontrollable events $\Sigma_u$, and a mask $P$. Similarly we use $\infpcC(K,L(G),\Sigma_u)$ to denote the infimal prefix-closed and controllable superlanguage and $\infpcO(K,L(G),P)$ to denote the infimal prefix-closed and observable superlanguage.

  Kumar and Shayman~\cite{KumarS98} have proved that the computation of the infimal prefix-closed, controllable and observable superlanguage of $K$ wrt $L(G)$ depends on the computation wrt $\Sigma^*$, namely
  $
    \infpcCO(K,L(G),\Sigma_u,P) = \infpcCO(K,\Sigma^*,\Sigma_u,P) \cap L(G)
  $.
  It thus suffices to consider the computation wrt the language $\Sigma^*$.
  They further proved that 
  $
    \infpcCO(K,\Sigma^*,\Sigma_u,P) = \infpcO (\infpcC(K,\Sigma^*,\Sigma_u),\Sigma^*,P)
  $.
  Lafortune and Chen~\cite{LC1990} have shown that $\infpcC(K,\Sigma^*,\Sigma_u) = \overline{K}\Sigma_u^*$, which can be computed from a DFA for $K$ in linear time. 
  The computation of the infimal prefix-closed and controllable superlanguage is thus easy and we focus in the rest on the computation of the infimal prefix-closed and observable superlanguage.
  
  Rudie and Wonham~\cite{RudieWonham1990} showed that $\infpcO(K,L(G),P) = L(G) \setminus (\Sigma^+ \setminus \widetilde{P}^{-1} (\widetilde{P}(\overline{K})))\Sigma^*$, where $P$ is a projection and $\widetilde{P}$ projects all but the last event, inductively defined by $\widetilde{P}(\eps) = \eps$ and $\widetilde{P}(sa) = P(s) a$. They also proved that for $K\neq\emptyset$,
  \begin{align}\label{formula1}
    \widetilde{P}^{-1} \widetilde{P}(\overline{K})
    = 
    \bigcup_{a\in\Sigma} \left[ P^{-1} (P(\overline{K}a \cap \overline{K}))\cap \Sigma^*a \right] 
    \cup \{\eps\}\,.
  \end{align}
  The equation remains valid for masks and Kumar and Shayman~\cite{KumarS98} extended it and simplified to the form
  \begin{align}\label{formula2}
    \infpcO(K,L(G),P) = \suppc[\widetilde{P}^{-1} \widetilde{P}(\overline{K})] \cap L(G)
  \end{align}
  where $\suppc(H)$ stands for the supremal prefix-closed sublanguage of a language $H$. Note that it immediately implies that $\infpcO(K,L(G),P) = \infpcO(K,\Sigma^*,P) \cap L(G)$.

  The formulae consist of operations studied in the literature and their worst-case state complexities give a rough estimate on the state complexity of the language $\infpcO(K,\Sigma^*,P)$. By Yu et al.~\cite{YuZS94}, the bound is no more than $2^{|\Sigma|(4n^2+8n+1)}$, where $n$ is the state complexity of $K$. Namely, Yu et al.~\cite{YuZS94} show that $\overline{K}a$ needs no more than $4n+8$ states and $\overline{K}a \cap \overline{K}$ no more than $(4n+8)n$ states. Then $P^{-1}P(\overline{K}a \cap \overline{K})$ needs at most $2^{(4n+8)n}$ states. (If $P$ is a natural projection, the bound is lower~\cite{JiraskovaM12,wong98}.) The intersection with $\Sigma^*a$ then needs no more than $2^{(4n+8)n} \cdot 2$ states and the union over all events $a$ in $\Sigma$ no more than $(2^{(4n+8)n} \cdot 2)^{|\Sigma|}$ states. The supremal prefix-closed sublanguage of a DFA can be computed in linear time and does not increase the state complexity; it requires to remove all non-marked states and corresponding transitions.
  
  Results of Yu et al.~\cite{YuZS94} hold for any language and the reader may notice that the languages of the formulae are of special forms. The worst-case state complexity of Yu et al.~\cite{YuZS94} is thus mostly not tight for them. For instance, it can be shown that the tight state complexity on $\overline{K}a\cap \overline{K}$ is $2n$ rather than $(4n+8)n$, which decreases the upper bound to $2^{|\Sigma|2n}$. 
  
  We now show that the upper bound on the state complexity of the language $\infpcO(K,\Sigma^*,P)$ is no more than $2^n + 1$. To this aim, we express the formula for $\infpcO(K,\Sigma^*,P)$ in an equivalent form using the operation of right quotient. This expression is based on the following relation between the mask, intersection and right quotient operations.
  \begin{lem}\label{lemma1}
    Let $P$ be a mask from $\Sigma$ to $\Delta$. For a prefix-closed language $K$ over $\Sigma$ and an event $a\in\Sigma$, it holds that 
    \[
      P^{-1} (P(Ka \cap K)) \cap \Sigma^*a = (P^{-1}P(K/a))a\,.
    \]
  \end{lem}
  \begin{IEEEproof}
    The claim holds for $K=\emptyset$. Assume that $K\neq\emptyset$.
    Let $xa \in P^{-1}(P(Ka\cap K))\cap \Sigma^*a$. Then $P(xa) \in P(Ka\cap K)$ and there exists $ya \in Ka\cap K$ such that $P(xa)=P(ya)$. Since $ya\in K$, we have that $y \in K/a$, hence $xa \in P^{-1}(P(y))a \subseteq (P^{-1}P(K/a))a$.
    On the other hand, let $xa \in (P^{-1}P(K/a))a$. Then $x\in P^{-1}P(K/a)$ and there is $y\in K/a$ with $P(x)=P(y)$. Since $y\in K/a$, $ya\in K$. Because $K$ is prefix-closed, $y\in K$, which implies that $ya \in Ka \cap K$. Thus, $P(xa)\in P(Ka\cap K)$, that is, $xa \in P^{-1}(P(Ka \cap K)) \cap \Sigma^*a$.
  \end{IEEEproof}

  The assumption that the language is prefix-closed is essential. The lemma does not hold for non-prefix-closed languages even if $P$ is the identity mask. In this case, Lemma~\ref{lemma1} reduces to $Ka \cap K = (K/a)a$. If $K=\{aa\}$ is non-prefix-closed, then $Ka\cap K=\emptyset$, whereas $(K/a)a=\{aa\}$.

  We can now express the formula of Kumar and Shayman~\cite{KumarS98} in an equivalent form using the operation of right quotient.
  
  \begin{thm}\label{thmNewFormula}
    Let $K$ be a nonempty language over $\Sigma$, and let $P$ be a mask from $\Sigma$ to $\Delta$. Then
    $
      \infpcO(K,\Sigma^*,P) = \suppc(\cup_{a\in\Sigma} (P^{-1} P(\overline{K}/a))a \cup \{\eps\}).
    $
  \end{thm}
  \begin{IEEEproof}
    By \eqref{formula1}, \eqref{formula2}, and Lemma~\ref{lemma1}, $\infpcO(K,\Sigma^*,P) = \suppc(\widetilde{P}^{-1} \widetilde{P}(\overline{K})) = \suppc ( \cup_{a\in\Sigma}\, [ P^{-1} (P(\overline{K}a \cap \overline{K})) \cap \Sigma^*a ] \allowbreak \cup \{\eps\} ) = \suppc ( \cup_{a\in\Sigma}\, (P^{-1} P(\overline{K}/a))a \cup \{\eps\} )$, respectively.
  \end{IEEEproof}

  We further modify the formula by moving the union operation deeper into the formula. It is then applied to a structurally simpler subformula, which is useful for our goal.
  
  \begin{lem}\label{LemDesc2}
    Let $K\subseteq \Sigma^*$ be a language and $P\colon \Sigma \to \Delta\cup\{\eps\}$ be a mask. Let $\Sigma'=\{a' \mid a\in\Sigma\}$ be a copy of $\Sigma$ disjoint from both $\Sigma$ and $\Delta$. Let $h\colon \Sigma\cup\Sigma' \to \Delta\cup\Sigma'\cup\{\eps\}$ be a mask defined by $h(a) = P(a)$, for $a\in\Sigma$, and $h(a')=a'$, for $a' \in \Sigma'$. Let $g\colon \Sigma' \to \Sigma$ be a mask defined by $g(a')=a$, for $a'\in\Sigma'$. Then
    \[
      \bigcup_{a\in\Sigma} (P^{-1}P(K/a)) a
      = 
      g \left( h^{-1}h \left( \bigcup_{a\in\Sigma} (K/a) a' \right) \cap \Sigma^*\Sigma' \right)\,.
    \]
  \end{lem}
  
  \begin{IEEEproof}
    By the properties of masks, we have that 
    \begin{align*}
      & \quad g ( h^{-1}(h ( \cup_{a\in\Sigma} (K/a) a' )) \cap \Sigma^*\Sigma' ) \\
      & = g ([\cup_{a\in\Sigma}\ h^{-1}(h((K/a) a'))] \cap \Sigma^*\Sigma' )\\
      & = g ([\cup_{a\in\Sigma}\ h^{-1}(h(K/a) h(a'))] \cap \Sigma^*\Sigma' )\\
      & = g ([\cup_{a\in\Sigma}\ h^{-1}(P(K/a) a')] \cap \Sigma^*\Sigma' )\\
      & = g ([\cup_{a\in\Sigma}\ h^{-1}(P(K/a)) h^{-1}(a')] \cap \Sigma^*\Sigma' )\\
      & = g ([\cup_{a\in\Sigma}\ P^{-1}(P(K/a)) a' P^{-1}(\eps)] \cap \Sigma^*\Sigma' )\\
      & = g (\cup_{a\in\Sigma}\ [P^{-1}(P(K/a)) a' P^{-1}(\eps) \cap \Sigma^*\Sigma' ])\\
      & = g (\cup_{a\in\Sigma}\ P^{-1}(P(K/a)) a')\\
      & = \cup_{a\in\Sigma}\ g(P^{-1}(P(K/a)) a')\\
      & = \cup_{a\in\Sigma}\ (P^{-1}P(K/a)) a\,.
    \end{align*}
    This completes the proof.
  \end{IEEEproof}

  As a corollary of Theorem~\ref{thmNewFormula} and Lemma~\ref{LemDesc2}, we obtain the following formula, which we use to show the asymptotically tight bound on the state complexity of $\infpcO(K,\Sigma^*,P)$.
  
  \begin{cor}\label{formulaComp}
    Under the assumptions of Lemma~\ref{LemDesc2}, if $K\neq\emptyset$,
    \begin{multline*}
      \infpcO(K,\Sigma^*,P) = \\ 
        \suppc \left[ 
          g \left( h^{-1}h \left( \bigcup_{a\in\Sigma} (\overline{K}/a) a' \right) \cap   \Sigma^*\Sigma' 
        \right) 
        \cup \{\eps\} \right] \,.
    \end{multline*}
  \end{cor}

\section{Deterministic State Complexity}
  We now use Corollary~\ref{formulaComp} to show that $2^n+1$ is an upper-bound on the state complexity of the language $\infpcO(K,\Sigma^*,P)$ and that the bound is asymptotically tight.

  Corollary~\ref{formulaComp} suggests an algorithm (Algorithm~\ref{alg1}) to compute the language $\infpcO(K,\Sigma^*,P)$.
  \begin{algorithm}[b]
    \caption{Computation of $\infpcO(K,\Sigma^*,P)$}
    \label{alg1}
    \begin{algorithmic}[1]
      \Require a DFA for $K$ over $\Sigma$ and a mask $P$
      \Ensure a DFA for the language $\infpcO(K,\Sigma^*,P)$
      \If{$K=\emptyset$}
        \Return the DFA for $K$
      \Else
      \State\label{step3} Compute a DFA for $\overline{K}$
      \State\label{step4} Compute a DFA for $\cup_{a\in\Sigma}\, (\overline{K}/a) a'$
      \State\label{step5} Compute an NFA for $g(h^{-1}h(\cup_{a\in\Sigma}\,(\overline{K}/a)a')\cap \Sigma^*\Sigma')$ 
      \State\label{step6} Determinize the NFA
      \State\label{step7} Compute the union with $\{\eps\}$
      \State\label{step8} Compute the supremal prefix-closed sublanguage
      \EndIf
    \end{algorithmic}
  \end{algorithm}
  We now discuss state complexities of its steps. Consequently we obtain its time complexity.
  \begin{lem}[Yu et al.~\cite{YuZS94}]\label{lemRQ}
    Let $\A$ be a DFA over $\Sigma$ with $n$ states, and let $a \in \Sigma$. Then the minimal DFA for $L_m(\A)/a$ has at most $n$ states. The bound is tight.
  \end{lem}
  
  The construction is as follows. Let $\A=(Q,\Sigma,\delta_{\A},q_0,F_{\A})$ be a DFA. Construct the DFA $\A'=(Q,\Sigma,\delta_{\A},q_0,F_{\A'})$, where $F_{\A'}=\{ q\in Q \mid \delta_{\A}(q,a) \in F_{\A}\}$. Then $L_m(\A') = L_m(\A)/a$. 
  
  We now study the size of the minimal DFA for the language computed in Step~\ref{step4} of the algorithm.
  \begin{lem}\label{lemma9}
    Let $\A$ be a DFA over $\Sigma$ with $n$ states. Then the minimal DFA for $\cup_{a\in\Sigma}\, (L_m(\A)/a)a'$ has at most $n+1$ states. The bound is tight even for prefix-closed languages.
  \end{lem}
  \begin{IEEEproof}
    Let $\A=(Q,\Sigma,\delta_{\A},q_0,F_{\A})$ be a DFA with $n$ states $Q=\{0,1,\ldots,n-1\}$. For every $a\in\Sigma$, we construct the set $F_{a} = \{ q \in Q \mid \delta_{\A}(q,a) \in F_{\A}\}$ of all states of $\A$ from which an $a$-transition reaches a marked state. We construct the DFA $\B=(Q\cup\{n\},\Sigma,\delta_{\B},0,\{n\})$ from $\A$ by adding a new state, $n$, which is the only marked state, and by defining the transitions $\delta_{\B}(q,a) = \delta_{\A}(q,a)$, for $0\le q \le n-1$ and $a\in\Sigma$, and $\delta_{\B}(f,a')=n$, for every $f\in F_{a}$. The construction is illustrated in Fig.~\ref{fig3}. The corresponding sets are $F_{a}=\{0,1\}$, $F_{b}=\{0\}$ and $F_{c}=\emptyset$.
    
    \begin{figure}
      \centering
      \begin{tikzpicture}[baseline,auto,->,>=stealth,shorten >=1pt,node distance=1.4cm,
        state/.style={ellipse,minimum size=6mm,very thin,draw=black,initial text=},
        every node/.style={font=\small}]
        \node[state,initial]    (1) {$0$};
        \node[state,accepting]  (2) [right of=1]  {$1$};
        \path
          (1) edge[bend left] node {$a,b$} (2)
          (2) edge[bend left] node[above] {$c$} (1)
          (2) edge[loop above] node {$a$} (2)
          ;
      \end{tikzpicture}
      \begin{tikzpicture}[baseline,auto,->,>=stealth,shorten >=1pt,node distance=1.4cm,
        state/.style={circle,minimum size=6mm,very thin,draw=black,initial text=},
        every node/.style={font=\small}]
        \node[state,initial]    (1) {$0$};
        \node[state]            (2) [right of=1] {$1$};
        \node[state,accepting]  (3) [right of=2] {$2$};
        \path
          (1) edge[bend left] node {$a,b$} (2)
          (2) edge[bend left] node[above] {$c$} (1)
          (2) edge[loop above] node {$a$} (2)
          (1) edge[bend right=40] node[above right] {$a',b'$} (3)
          (2) edge[bend left=20] node {$a'$} (3)
          ;
      \end{tikzpicture}
      \caption{Automata $\A$ (left) and $\B$ (right) for $\cup_{a\in\Sigma} (L_m(\A)/a)a'$}
      \label{fig3}
    \end{figure}
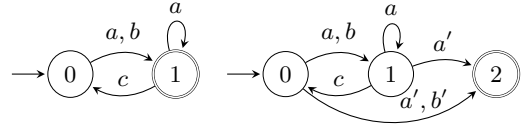
    
    We claim that $\B$ marks the language $\cup_{a\in\Sigma}\, (L_m(\A)/a)a'$. If a string is marked by $\B$, it is of the form $wa'$, for some $a\in\Sigma$, which means that $\delta_{\B}(0,w) \in F_{a}$. By the construction of $F_{a}$, $w \in L_m(\A)/a$, hence $wa' \in (L_m(\A)/a)a'$.
    On the other hand, if $wa' \in \cup_{a\in\Sigma}\,(L_m(\A)/a)a'$, then $w\in L_m(\A)/a$, hence $\delta_{\B}(0,w)=f_{a}$, for some $f_{a} \in F_{a}$, which implies that $\delta_{\B}(0,wa') = \delta_{\B}(f_{a},a') = n$, hence it is marked by $\B$.

    To show that the bound is tight, we consider the DFA $\A$ depicted in Fig.~\ref{fig1} (solid arrows) with states $\{0,\ldots,n-1\}$, where state $0$ is initial and all states are marked. The DFA is minimal; two states are distinguishable by a string in $b^*$. The DFA $\B$ for $(L_m(\A)/a)a' \cup (L_m(\A)/b)b'$ is depicted in Fig.~\ref{fig1} (all arrows), where the states are $\{0,\ldots,n\}$ with $n$ being the only marked state. There is an $a'$-transition from state $i$ to state $n$ for every $0\le i\le n-1$, and a $b'$-transition from state $j$ to state $n$ for every $0\le j\le n-2$. The DFA $\B$ is minimal; states $\{0,\ldots,n-1\}$ are distinguishable by the same argument as for $\A$ and $n$ is not equivalent with any other state since it is the only marked state.
  \end{IEEEproof}

  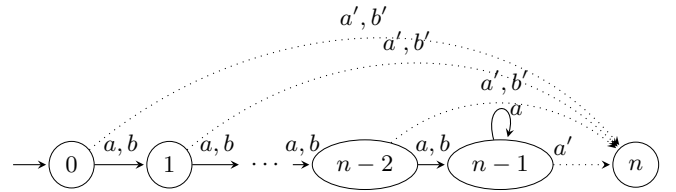
\begin{figure}[b]
    \centering
    \begin{tikzpicture}[baseline,auto,->,>=stealth,shorten >=1pt,node distance=1.3cm,
      state/.style={ellipse,minimum size=6mm,very thin,draw=black,initial text=},
      every node/.style={font=\small}]
      \node[state,initial]  (1) {$0$};
      \node[state]          (4) [right of=1]  {$1$};
      \node[]               (5) [right of=4]  {$\cdots$};
      \node[state]          (4') [right of=5] {$n-2$};
      \node[state]          (5') [right of=4',node distance=1.8cm]  {$n-1$};
      \node[state]          (s) [right of=5',node distance=1.8cm]  {$n$};
      \path
        (1) edge node {$a,b$} (4)
        (4) edge node {$a,b$} (5)
        (5) edge node {$a,b$} (4')
        (5') edge[loop above] node[right] {$a$} (5')
        (4') edge node {$a,b$} (5')
        (1) edge[dotted,bend right=0,out=40,in=130] node {$a',b'$} (s)
        (5') edge[dotted] node[above left] {$a'$} (s)
        (4') edge[dotted,out=40,in=140] node {$a',b'$} (s)
        (4) edge[dotted,out=40,in=135] node {$a',b'$} (s)
        ;
    \end{tikzpicture}
    \caption{Automata $\A$ (solid arrows) and $\B$ (all arrows)}
    \label{fig1}
  \end{figure}

  We now use the previous results to obtain our upper-bound on the state complexity of the language $\infpcO(K,\Sigma^*,P)$.

  \begin{thm}[Upper bound]\label{thmUpperBound}
    Let $K$ over $\Sigma$ be a nonempty language marked by a DFA with $n$ states. Then the minimal DFA for $\infpcO(K,\Sigma^*,P)$ has no more than $2^{n} + 1$ states.
  \end{thm}
  \begin{IEEEproof}
    Let $P\colon \Sigma\to \Delta\cup\{\eps\}$. By Corollary~\ref{formulaComp}, we have that $\infpcO(K,\Sigma^*,P) = 
        \suppc [ 
          g( h^{-1}h ( \cup_{a\in\Sigma}\, (\overline{K}/a) a' ) \cap \Sigma^*\Sigma' )
        \cup \{\eps\} ]$.
    From Lemma~\ref{lemma9}, we have that the minimal DFA marking the language $\cup_{a\in\Sigma}\, (\overline{K}/a) a'$ has at most $n+1$ states, only one of which is marked. We denote this state by $f$. Notice that, by the construction, there is no transition from state $f$. 
    
    We represent the language $g( h^{-1}h ( \cup_{a\in\Sigma}\, (\overline{K}/a) a' ) \cap \Sigma^*\Sigma' )$ as an NFA as follows. The language $h(\cup_{a\in\Sigma}\, (\overline{K}/a) a')$ is computed by replacing every $x$-transition, $x\in\Sigma$, with the $h(x)$-transition. The language $h^{-1}h(\cup_{a\in\Sigma}\, (\overline{K}/a) a')$ is then computed by replacing every $y$-transition, $y\in\Delta$, by an $x$-transition for every $x\in\Sigma$ such that $h(x)=y$. In addition, for every $x\in\Sigma$ such that $h(x)=\eps$, we add a self-loop under $x$ to every state of the NFA. To compute an NFA for $h^{-1}h(\cup_{a\in\Sigma}\, (\overline{K}/a) a') \cap \Sigma^*\Sigma'$ then means to remove all transitions from state $f$. This can be done during the computation of an NFA for $h^{-1}h(\cup_{a\in\Sigma} (\overline{K}/a) a')$ so that no self-loop is added to state $f$. The computation of an NFA for the mask $g$ is similar to that of $h$. 
    
    The resulting NFA has at most $n+1$ states. Thus, a DFA equivalent to the NFA, constructed by the standard subset construction, has at most $2^{n+1}$ reachable states. However, since every marked state of the subset automaton must contain $f$, and there are at most $2^{n}$ subsets containing $f$, there are at most $2^n$ marked states in the computed DFA. 

    To compute the union with $\{\eps\}$, the DFA may require one more (initial and marked) state. Thus, the resulting DFA has at most $2^{n+1} + 1$ states, where at most $2^n + 1$ states are marked.

    Since $\infpcO(K,\Sigma^*,P)$ is prefix-closed, its minimal DFA must have all states marked. There are at most $2^n + 1$ marked states in the above constructed automaton, therefore the minimal DFA for $\infpcO(K,\Sigma^*,P)$ can have at most so many states.
  \end{IEEEproof}

  Consequently, the time complexity of Algorithm~\ref{alg1} is $O(2^n)$. Indeed, let $n$ be the state complexity of $K$. Step~\ref{step3} requires time $O(n)$. To compute Step~\ref{step4}, we add a new state, $f$, and scan the automaton in linear time using, e.g., the breadth-first search (BFS) algorithm~\cite{introToAlgs}. For every state $q$ and its out-going transition under $x$, if $\delta(q,x)$ is marked, we add an $x'$-transition from $q$ to $f$. This can be done in time $O(1+n+2n\cdot |\Sigma|) = O(n\cdot |\Sigma|)$, since there are $n$ states, one added new state, and at most $n\cdot |\Sigma|$ transitions that may be duplicated to $f$. Step~\ref{step5} can be computed in time $O(n\cdot |\Sigma|)$ as follows. The application of $h$ can be done in time $O(n\cdot |\Sigma|)$ by the BFS algorithm. The application of $h^{-1}$ can be done in time $O(n\cdot |\Sigma| + n\cdot |\Sigma\setminus\Delta|) = O(n\cdot |\Sigma|)$, where the second part corresponds to adding self-loops under unobservable events. As explained above, the intersection with $\Sigma^*\Sigma'$ is done so that no transitions are added to $f$ during the computation of $h^{-1}$. Step~\ref{step6} can be computed in time $O(2^n \cdot |\Sigma|)$, since, by the proof of Theorem~\ref{thmUpperBound}, the DFA has at most $2^{n+1}+1$ states and $|\Sigma|$ transitions in every state. Step~\ref{step7} can be computed in time $O(|\Sigma|)$ as follows: let $q_0$ be the initial state of the DFA, and let $q_i$ be a new marked state. We change the DFA so that $q_i$ is the only initial state, i.e., $q_0$ is not initial anymore, and for every $x\in\Sigma$, we define $\delta(q_i,x)=\delta(q_0,x)$. Finally, Step~\ref{step8} can be computed in linear time wrt the size of the input DFA by removing all non-marked states and the corresponding transitions. The overall time complexity is $O(|\Sigma|\cdot 2^n)$. Considering the size of the alphabet as constant results in the claimed complexity $O(2^n)$.
  
  We now discuss the lower-bound state complexity and show that it is $\Omega(2^{n})$. It holds even for projections.
  
  \begin{thm}[Lower bound]\label{thm7b}\label{thmLowerBound}
    Let $P\colon \{a,b,c\}^* \to \{a,b\}^*$ be a projection. For every $n\ge 2$, there exists a minimal DFA with $n$ states marking a language $K_n\subseteq \{a,b,c\}^*$, such that the state complexity of $\infpcO(K_n,\Sigma^*,P)$ is at least $\frac{3}{4}\cdot 2^{n}-1$.
  \end{thm}
  \begin{IEEEproof}
    Let $K_n$ be the language marked by the DFA $\A_n$ depicted in Fig.~\ref{fig:thm5a}.
    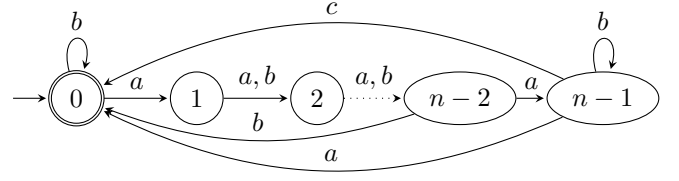
\begin{figure}
      \centering
      \begin{tikzpicture}[->,>=stealth,shorten >=1pt,auto,node distance=1.6cm,
      state/.style={ellipse,minimum size=7mm,draw=black,initial text=}]

      \node[state,initial,accepting]   (0) {$0$};
      \node[state] (1) [right of=0] {$1$};
      \node[state] (a) [right of=1] {$2$};
      \node[state] (2) [right of=a,node distance=1.9cm] {$n-2$};
      \node[state] (3) [right of=2,node distance=1.9cm] {$n-1$};

      \path
        (0) edge node {$a$} (1)
        (0) edge[loop above] node[above] {$b$} (0)
        (1) edge node {$a,b$} (a)
        (2) edge node {$a$} (3)
        (a) edge[dotted] node {$a,b$} (2)
        (3) edge[loop above] node[above] {$b$} (3)
        (3) edge[bend left=25] node[above] {$a$} (0)
        (3) edge[bend right=26] node[above] {$c$} (0)
        (2) edge[bend left=19] node[above] {$b$} (0)
        ;
      \end{tikzpicture}
      \caption{The minimal DFA $\A_n$ for $K_n$}
      \label{fig:thm5a}
    \end{figure}
    It has $n$ states $\{0,1,\ldots,n-1\}$, where state $0$ is the sole initial and marked state. For $0\le i\le n-1$, $\delta(i,a)= (i+1 \bmod n)$. For $1\le i\le n-3$, $\delta(i,b)= i+1$, $\delta(n-2,b)=0$, and, for $i\in\{0,n-1\}$, $\delta(i,b)=i$. Finally, there is a single $c$-transition $\delta(n-1,c)=0$.

    An NFA $\B_n$ for the language $g(h^{-1}h ( \cup_{a\in\Sigma} (\overline{K_n}/a) a' ) \cap \Sigma^*\Sigma')$ is build from $\A_n$ according to the above constructions in the following steps and the result is depicted in Fig.~\ref{fig:thm5b}:
    \begin{enumerate}
      \item We compute $\overline{K_n}$ by marking all states of $\A_n$.
      \item To compute $\cup_{a\in\Sigma} (\overline{K_n}/a)a'$, we add a new state, $n$. From every state of $\A_n$, transitions under $a'$ and $b'$ go to state $n$, and a transition under $c'$ goes from state $n-1$ to state $n$. The only marked state is state $n$.
      \item The language $h(\cup_{a\in\Sigma} (\overline{K}/a)a')$ is computed by replacing the $c$-transition by an $\eps$-transition.
      \item To compute $h^{-1}h(\cup_{a\in\Sigma} (\overline{K}/a)a') \cap \Sigma^* \Sigma'$, a self-loop under $c$ is added to every state of $\A_n$. Note that it is not added to state $n$, since it would be eliminated by the intersection with $\Sigma^* \Sigma'$. Thus, this can be done in linear time without computing the intersection.
      \item Finally, to apply $g$ means to rename all transitions under $a'$, $b'$ and $c'$, which all go to state $n$.
    \end{enumerate}
    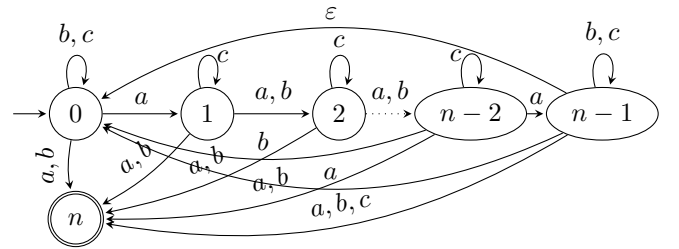
\begin{figure}[b]
      \centering
      \begin{tikzpicture}[->,>=stealth,shorten >=1pt,auto,node distance=1.75cm,
        state/.style={ellipse,minimum size=7mm,draw=black,initial text=}]

        \node[state,initial]   (0) {$0$};
        \node[state,accepting] (4) [below of=0,node distance=1.4cm] {$n$};
        \node[state] (1) [right of=0] {$1$};
        \node[state] (a) [right of=1] {$2$};
        \node[state] (2) [right of=a] {$n-2$};
        \node[state] (3) [right of=2] {$n-1$};

        \path[every node/.style={sloped,anchor=south,auto=false}]
          (0) edge node {$a$} (1)
          (0) edge[loop above] node[above] {$b,c$} (0)
          (1) edge node {$a,b$} (a)
          (2) edge node {$a$} (3)
          (a) edge[dotted] node {$a,b$} (2)
          (3) edge[loop above] node[above] {$b,c$} (3)
          (3) edge[bend left=26] node[above] {$a$} (0)
          (2) edge[bend left=20] node[above] {$b$} (0)
          (3) edge[bend right=30] node[above] {$\eps$} (0)
          (0) edge[bend right=10] node {$a,b$} (4)
          (1) edge[bend left=10] node {$a,b$} (4)
          (a) edge[bend left=10] node {$a,b$} (4)
          (2) edge[bend left=15] node[above] {$a,b$} (4)
          (3) edge[bend left=20] node[above] {$a,b,c$} (4)
          (1) edge[loop above] node[right] {$c$} (1)
          (a) edge[loop above] node {$c$} (a)
          (2) edge[loop above] node[left] {$c$} (2)
          ;
      \end{tikzpicture}
      \caption{An NFA $\B_n$ marking language $g(h^{-1}h \left( \bigcup_{a\in\Sigma} (\overline{K_n}/a) a' \right) \cap \Sigma^*\Sigma')$}
      \label{fig:thm5b}
    \end{figure}
    
    We show that the minimal DFA equivalent to the NFA $\B_n$ has at least $\frac{3}{4}\cdot 2^{n} - 1$ reachable marked states. Using the standard subset construction, we first show that all states of the subset automaton corresponding to the NFA $\B_n$ are pairwise distinguishable. Indeed, $\B_n$ marks $\eps$ only from state $n$ and $a^i c$ only from state $n-1-i$, for $0\le i\le n-1$. Therefore, the states of the subset automaton are pairwise distinguishable. To prove the theorem, we show that the subset automaton has $2^{n-1}+ 2^{n-2} - 1$ marked states that are all reachable via other marked states.

    State $\{0\}$ is initial, but not marked; we resolve this issue later. We now prove, by induction on the size of the subset, that every subset of $\{0,1,\ldots,n-1,n\}$ containing 0 and $n$ is reachable in the subset automaton from state $\{0\}$ by a nonempty string over $\{a,b\}$. Since there is an $a$-transition and a $b$-transition from every state $0$ through $n-1$ to $n$, all subsets reachable by such a string must contain state $n$, i.e., they are marked in the subset automaton. State $\{0,n\}$ is reachable from state $\{0\}$ by $b$. State $\{n-2,n\}$ is reachable from $\{0\}$ by $a^{n-2}$. State $\{0,n-2,n\}$ is reachable from state $\{n-2,n\}$ by $a^2b^{n-3}$. State $\{0,n-2,n\}$ goes to state $\{0,1,n-1,n\}$ by $a$, and then by a string in $b^*$ to states $\{0,i,n-1,n\}$ with $1\le i \le n-2$. State $\{0,n-2,n-1,n\}$ goes to state $\{0,n-1,n\}$ by $b$, and then to state $\{0,1,n\}$ by $a$. By a string in $b^*$, state $\{0,1,n\}$ goes to states $\{0,i,n\}$ with $1\le  i\le n-2$. Thus, each subset of size two or three containing 0 and $n$ is reachable.
 
    Now, let $X=\{0,i_1,i_2,\ldots,i_t,n\}$ be a set of size $t+2$, where $2\le t \le n-1$ and $1\le i_1<i_2<\cdots <i_t\le n-1$. We consider two cases:
    \begin{enumerate}
    
      \item If $i_t=n-1$, then $X$ is reachable from state $\{0,i_2-i_1,\ldots,i_{t-1}-i_1,n-2,n\}$ by $ab^{i_1-1}$, and the latter set of size $t+1$ is reachable by the induction hypothesis.
    
      \item If $i_t<n-1$, then $X$ is reachable from state $\{0,i_2-i_1,\ldots,i_t-i_1,n-1,n\}$ by $ab^{i_1-1}$, and the latter set of size $t+2$ contains state $n-1$, and is reachable by 1).
    
    \end{enumerate}
    This proves reachability of all subsets of $\{0,1,\ldots,n\}$ containing 0 and $n$. There are $2^{n-1}$ such subsets.
 
    Next, if $X=\{i_1,i_2,\ldots,i_t\}$ is a non-empty subset of the set $\{1,2,\ldots,n-2\}$, then the set $X\cup\{n\}$ is reachable from the set $\{0,i_2-i_1,i_3-i_1,\ldots,i_t-i_1,n\}$ containing 0 and $n$ by $a^{i_1}$. Thus, for every $\emptyset \neq X \subseteq \{1,2,\ldots,n-2\}$, state $X\cup\{n\}$ is reachable in the subset automaton. These sets do not contain $0$, hence they are different from the reachable states considered above. There are $2^{n-2}-1$ such subsets.
  
    Finally, we compute the union with the language $\{\eps\}$. To do this, we create a new initial and accepting state, $I$, (state $\{0\}$ is not initial anymore) with transitions defined exactly as for state $\{0\}$, that is, $\delta(I,x)=\delta(\{0\},x)$, for every $x\in\{a,b,c\}$. This has resolved the problem with the non-marked initial state, since state $I$ is marked and has the same transitions as state $\{0\}$, that is, all states reachable from state $\{0\}$ are also reachable from state $I$. Thus, we have shown that the minimal DFA constructed by the subset construction has at least $2^{n-1}+2^{n-2}$ marked states that are all reachable from the initial marked state $I$ via marked states.
    
    However, state $I$ is equivalent to state $\{0,n\}$. Indeed, both states $I$ and $\{0,n\}$ go to state $\{1,n\}$ under $a$, to state $\{0,n\}$ under $b$, and to state $\{0\}$ under $c$. 
    
    It remains to show that if the non-marked states are eliminated, the constructed marked states different from $I$ are still pairwise distinguishable. Let $X$ and $Y$ be two sets different from $I$ constructed above. They both contain $n$ and, without loss of generality, we may assume that there exists $i$ such that $n-1-i \in X \setminus Y$. Then the set reachable from $X$ under $a^{i}$ contains $n-1$, but the set reachable from $Y$ under $a^{i}$ does not. It means that $a^ic$ is marked from $X$, but not from $Y$, which distinguishes the states $X$ and $Y$. Therefore, the minimal DFA of the supremal prefix-closed sublanguage has at least $2^{n-1} + 2^{n-2} - 1$ states, which completes the proof.
  \end{IEEEproof}
  
  Combining the upper and lower bounds of Theorems~\ref{thmUpperBound} and~\ref{thmLowerBound} gives the following corollary.
  \begin{cor}
    Let $K$ over $\Sigma$ be a language with state complexity $n$, and let $P$ be a mask. Then the worst-case state complexity of the language $\infpcO(K,\Sigma^*,P)$ is $\Theta(2^n)$. 
    \hfill\IEEEQEDclosed
  \end{cor}

  We also have the following consequence on the time complexity of Algorithm~\ref{alg1}.
  \begin{cor}
    The time complexity of Algorithm~\ref{alg1} is $\Theta(2^n)$, where $n$ is the state complexity of the input language. \hfill\IEEEQEDclosed
  \end{cor}

\section{Nondeterministic State Complexity}\label{complTheory}
  Algorithm~\ref{alg1} represents the language as an NFA and it is determinized before computing the operation $\suppc(\cdot)$. The algorithm computing $\suppc(\cdot)$ on a DFA cuts off all non-marked states and the corresponding transitions, which requires linear time wrt the size of the input DFA. However, as shown above, this DFA may be exponentially larger than the DFA for $K$. 
  
  Another possibility is to execute $\suppc(\cdot)$ directly on an NFA. We now discuss this possibility and show that, in general, there is no polynomial-time algorithm that, given an NFA $\A$, would compute an NFA marking the language $\suppc(L_m(\A))$. 
  
\begin{figure}
    \centering
    \begin{tikzpicture}[baseline,auto,->,>=stealth,shorten >=1pt,node distance=1.4cm,
      state/.style={ellipse,minimum size=0mm,very thin,draw=black,initial text=},
      every node/.style={font=\small}]
      \node[state,initial,accepting]  (1) {$0$};
      \node[state]                    (2) [right of=1] {$1$};
      \node[state,accepting]          (3) [right of=2] {$2$};
      \path
        (1) edge node {$a$} (2)
        (1) edge[bend right=40] node[above right] {$a$} (3)
        (2) edge node {$b$} (3)
        ;
    \end{tikzpicture}
    \caption{A prefix-closed NFA $\A$ with $L_m(\A)=\overline{\{ab\}}$}
    \label{fig4}
  \end{figure}
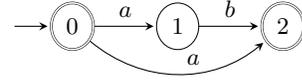

  We first provide a brief insight into the difference between the computation of $\suppc(\cdot)$ for DFAs and NFAs. Indeed, if all states of an NFA are marked, then its language is prefix-closed. However, compared to DFAs, the problem with NFAs is that having a non-marked state does not yet mean that the language is not prefix-closed, cf. Fig.~\ref{fig4} for an example. It can be shown that, given an NFA, it is PSPACE-complete to decide whether its marked language is prefix-closed~\cite{TM2016}.
  
  \begin{thm}\label{thmCompPC}
    The problem whether the marked language of an NFA is prefix-closed is PSPACE-complete.
    \hfill\IEEEQEDclosed
  \end{thm}

  We now show that there is no polynomial-time algorithm computing an NFA representation of $\suppc(L_m(\A))$ in general.
  
  \begin{thm}
    Let $\A$ be an NFA. There is no polynomial-time algorithm computing an NFA for the language $\suppc(L_m(\A))$. The claim holds even for unary languages.
  \end{thm}
  \begin{IEEEproof}
    We prove the theorem by constructing, for any $n\ge 1$, an NFA $\A_n$ with polynomially many states in $n$ such that any NFA for $\suppc(L_m(\A_n))$ has at least exponentially many states in $n$. Clearly, such an NFA cannot be computed in polynomial time wrt the size of $\A_n$.
    
    To construct the NFAs $\A_n$, we first construct auxiliary DFAs $\B_n$, for every $n\ge 0$. The DFA $\B_0=(X_0, \allowbreak \{a\}, \allowbreak \gamma_0, \allowbreak X_{i,0}, \allowbreak X_{m,0})$, where $X_0=X_{i,0}=X_{m,0}=\{0_0\}$ and $\gamma_0(0_0,a)$ is undefined. For $n\ge 1$, let $p_n$ denote the $n$th prime number. We define the DFA $\B_n=(X_n,\{a\},\gamma_n,X_{i,n},X_{m,n})$, where the state set is $X_n=\{0_n,1_n,\ldots,(p_n-1)_n\}$, the set of initial states is $X_{i,n}=\{0_n\}$, the set of marked states is $X_{m,n}=X_n\setminus\{0_n\}$, and the transition function is $\gamma_n(i_n,a)=(i+1 \bmod p_n)_n$, for all $i_n\in X_n$. Then $L_m(\B_0)=\{\eps\}$ and $L_m(\B_n)=a^* \setminus (a^{p_n})^*$, cf. Fig.~\ref{fig5} for automata $\B_0$, $\B_1$, and $\B_2$. We assume that the state sets $X_i$ and $X_j$ are disjoint for any $i\neq j$.
    
    For $n\ge 1$, we build the NFA $\A_n = (Q_n,\{a\},\delta_n,Q_{i,n},F_n)$ as a ``nondeterministic'' union of the DFAs $\B_0,\B_1,\ldots,\B_n$. The NFA $\A_2$ is depicted in Fig.~\ref{fig5}. Formally,
      $Q_n = \cup_{k=0}^{n}\, X_k$,
      $\delta_n(i_k,a) = \gamma_k(i_k,a)$,
      $Q_{i,n} = \cup_{k=0}^{n}\, X_{i,k}$, and
      $F_n = \cup_{k=0}^{n}\, X_{m,k}$.
    \begin{figure}
      \centering
      \begin{tikzpicture}[baseline,auto,->,>=stealth,shorten >=1pt,node distance=2cm,
        state/.style={ellipse,minimum size=0mm,very thin,draw=black,initial text=},
        every node/.style={font=\small}]
        \node[state,initial,accepting]  (00) {{\tiny $0_0$}};
        \node[state,initial]    (11) [right of=00,node distance=2cm] {{\tiny $0_1$}};
        \node[state,accepting]  (21) [right of=11] {{\tiny $1_1$}};
        \path
          (11) edge node {$a$} (21)
          (21) edge[bend left=40] node[above] {$a$} (11)
          ;
          
        \node[state,initial]    (12) [below of=00,node distance=1cm] {{\tiny $0_2$}};
        \node[state,accepting]  (22) [right of=12] {{\tiny $1_2$}};
        \node[state,accepting]  (32) [right of=22] {{\tiny $2_2$}};
        \path
          (12) edge node {$a$} (22)
          (22) edge node {$a$} (32)
          (32) edge[bend left=33] node[above] {$a$} (12)
          ;
      \end{tikzpicture}
      \caption{The NFA $\A_2$; ``nondeterministic'' union of DFAs $\B_0$, $\B_1$, and $\B_2$}
      \label{fig5}
    \end{figure}
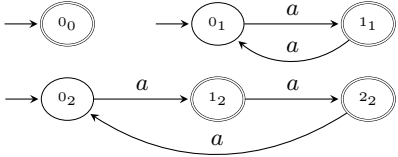
    The number of states of $\A_n$ is $1+\sum_{i=1}^{n} p_i$, which has been estimated by Bach and Shallit~\cite{BachShallit1996} to be $1+2^{-1} n^2 \ln n = O(n^2 \ln n)$.
    The marked language of $\A_n$ is $L_m(\A_n)= a^* \setminus (a^{p_n\#})^+$, where $p_n\# = \Pi_{i=1}^{n} p_i$. Indeed, for $m\ge 1$, string $a^m$ is marked by $\A_n$ if and only if there is $p_i\in\{p_1,\ldots,p_n\}$ such that $m \bmod p_i \neq 0$. Thus, the shortest string that is not marked by $\A_n$ is of length $p_n\#$. Therefore, the supremal prefix-closed sublanguage of $L_m(\A_n)$ is the finite language $\overline{\{a^{p_n\#-1}\}}$. 
    
    We now show, using the fooling set technique~\cite{Birget1992}, that any NFA marking this language requires at least $p_n\#$ states. 
    \begin{fact}[Fooling set technique]\label{fst}
      Let $L\subseteq \Sigma^*$ be a language, and let $S=\{(x_i,y_i) \mid 1\le i\le k\}$ be a set of pairs such that
      \begin{itemize}
        \item[(i)] $x_iy_i\in L$ for $1\le i \le k$, and
        \item[(ii)] if $i\neq j$, then $x_iy_j\notin L$ or $x_jy_i \notin L$, for $1\le i,j\le k$.
      \end{itemize}
      Then any NFA marking the language $L$ has at least $k$ states. Set $S$ is called a fooling set for $L$.\hfill\IEEEQEDopen
    \end{fact}

    Let $S=\{(a^i,a^{p_n\#-i-1}) \mid 0\le i\le p_n\#-1\}$. Then $a^{i+p_n\#-i-1}=a^{p_n\#-1}$ belongs to the language $\overline{\{a^{p_n\#-1}\}}$. Thus, $S$ satisfies item (i) of the fooling set technique. To show that it also satisfies item (ii), let $(a^i,a^{p_n\#-i-1})$ and $(a^j,a^{p_n\#-j-1})$ be two elements of $S$. Without loss of generality, we assume that $i<j$. Then $j + p_n\#-i-1 > p_n\#-1$, which implies that $a^{j}a^{p_n\#-i-1}$ does not belong to $\overline{\{a^{p_n\#-1}\}}$, i.e., it proves that $S$ satisfies item (ii). Thus, $S$ is a fooling set for the language $\overline{\{a^{p_n\#-1}\}}$ of size $p_n\#$. Therefore, any NFA marking the language $\overline{\{a^{p_n\#-1}\}}$ has at least $p_n\#$ states. 
    Since $p_n\# = e^{(1+o(1))n\log n}$~\cite{oeis} is exponential wrt $n$, hence not polynomial wrt the size of $\A_n$, there is no algorithm that would compute an NFA for the language $\overline{\{a^{p_n\#-1}\}}$ in polynomial time.
  \end{IEEEproof}

\section{Conclusion}
  A consequence of the exponential state complexity is that any algorithm computing a DFA for $\infpcO(K,\Sigma^*,P)$ requires, in the worst case, exponential time (and exponential space to store it). Algorithm~\ref{alg1} further shows that the exponential time is sufficient. The algorithm is thus optimal in the sense that there is no asymptotically more efficient algorithm.
  
  Concerning the NFA representation, we showed that even for unary languages, the algorithm would need more than polynomial time to compute the result and more than polynomial space to store it. This is in contrast to checking whether the language of an NFA is prefix closed, which can be done in polynomial space and it is not known whether it can be done in polynomial time.

\section*{Acknowledgment}
  The author gratefully acknowledges very useful suggestions and comments of the anonymous referees. The upper bound on the state complexity of the operation $\overline{K}a\cap\overline{K}$ is due to an anonymous referee.

\bibliographystyle{IEEEtran}
\bibliography{mybib}

\end{document}